\documentclass[%
 reprint, 
 amsmath,amssymb,
 aps,
 nofootinbib
 prb,
 superscriptaddress,
 citeautoscript
 ]{revtex4-2}
\usepackage{scrextend}
\usepackage{graphicx}
\usepackage{dcolumn}
\usepackage{bm}
\usepackage{tabularx}
\usepackage[utf8]{inputenc}
\usepackage[english]{babel}
\usepackage{color}

\usepackage{physics}

\usepackage{amsmath}
\usepackage{soul}
\DeclareMathAlphabet{\pazocal}{OMS}{zplm}{m}{n}

\usepackage{hyperref}
\usepackage{comment}

\definecolor{mycyan}{rgb}{0, 1, 1}

\definecolor{mygreen}{rgb}{0, 1, 0}

\definecolor{mypink}{rgb}{1, 0.753, 0.796}

\begin{document}

\preprint{APS/123-QED}

\title{Charge dynamics at nitrogen impurities and nitrogen-vacancy centers in diamond}

\author{Chandan Kumar Vishwakarma}
\affiliation{Materials Department, University of California, Santa Barbara, California 93106-5050, USA}
\author{J. K. Nangoi}
\affiliation{Materials Department, University of California, Santa Barbara, California 93106-5050, USA}

\author{Mark E. Turiansky}
\affiliation{Materials Department, University of California, Santa Barbara, California 93106-5050, USA}
\affiliation{US Naval Research Laboratory, 4555 Overlook Avenue SW, Washington, DC 20375, USA}

\author{Chris G. Van de Walle}
\email{vandewalle@mrl.ucsb.edu}
\affiliation{Materials Department, University of California, Santa Barbara, California 93106-5050, USA}


             
\begin{abstract}
The nitrogen-vacancy (NV) center in diamond is the prototype quantum defect that enables a variety of diamond-based quantum technologies. However, charge-state instability and spectral diffusion, often induced by substitutional nitrogen impurities (N$_{\rm C}$), remain key challenges for device performance. 
Here, we employ first-principles density functional theory calculations to quantitatively investigate nonradiative carrier capture processes mediated by multiphonon emission at both the NV center and the N$_{\rm C}$ impurity. 
For relevant cases, we also compute the rates of radiative and thermal emission processes. 
For N$_{\rm C}^0$ $\to$ N$_{\rm C}^-$, we obtain an electron capture coefficient of $2.2 \times 10^{-8}$~cm$^3$~s$^{-1}$ at 300~K.
Both the magnitude and temperature dependence are in excellent agreement with experimentally measured capture cross sections.
Electron capture at N$_{\rm C}^+$ is even faster, with a capture coefficient of 
$1.0 \times 10^{-4}$~cm$^3$~s$^{-1}$ at 300~K.
For the NV center, we find that carrier capture rates involving only the ground states of NV$^0$ and NV$^-$ are negligibly slow.
However, capture into the \textit{excited} states (NV$^{0*}$ and NV$^{-*}$) is significantly faster.
In particular, the capture coefficient for the hole capture process NV$^-$ $\to$ NV$^{0*}$ is as large as 
$1.8 \times 10^{-7}$~cm$^3$~s$^{-1}$ 
and largely temperature-independent.
Hole capture at NV$^-$ will thus occur via nonradiative capture into an excited state of NV$^{0}$ followed by fast radiative decay to the NV$^0$ ground state. 
Similarly, electron capture at NV$^0$ will occur via the NV$^0$ $\to$ NV$^{-*}$ $\to$ NV$^-$ pathway, but with a lower nonradiative capture coefficient 
($2.1 \times 10^{-9}$~cm$^3$~s$^{-1}$ at 300~K). 
Our calculated capture coefficients and rates provide essential information for analyzing charge-state dynamics.

\end{abstract}

\maketitle

\section{Introduction} \label{sec:intro}

Quantum technologies are being developed for computation, communication, and sensing,
and point defects in wide-band-gap semiconductors have emerged as a promising platform due to their robustness and potential for scalability and integration~\cite{Weber-2010, Bassett-2019, Mark-2024}. 
The nitrogen-vacancy (NV) center in diamond has received notable attention as a prototype solid-state qubit \cite{Doherty-2013,Awschalom-2018}.  In the negative charge state (NV$^-$), it offers optically addressable spin states that can be initialized, manipulated, and read out at room temperature  \cite{Gruber-1997, Siyushev-2019}, making it a leading platform for quantum information, and also quantum sensing including nanoscale magnetometry \cite{Rovny-2024, Smith-2018}

The unique properties of NV$^-$ arise from its electronic structure. 
Three spin-split defect states reside within the diamond band gap and are occupied by four electrons{~\cite{Lenef-1996}}. 
In the ground state, the electrons form a triplet configuration with $^3A_2$ symmetry (within the $C_{3v}$ point group). An excited state with $^3E$ symmetry 
is accessible through a spin-conserving optical transition with a zero-phonon line (ZPL) at 1.945~eV (637~nm) \cite{Doherty-2013}. 

The presence of substitutional nitrogen impurities (N$_{\rm C}$) plays a critical role in the formation and properties of NV centers. 
Nitrogen is of course essential for the creation of the NV center, but it also influences its operation and charge-state stability. 
Optical excitation may ionize substitutional nitrogen impurities (N$_{\rm C}$) since their (+/0) charge-state transition level is 1.73~eV from the conduction-band minimum (CBM) \cite{Farrer-1969, Koppitz-1986}. 
N$_{\rm C}^+$ impurities that are sufficiently close to NV$^-$ can induce a Stark shift that leads to spectral diffusion~\cite{Manson-2018}. 
Charge transfer between and N$_{\rm C}$ and NV can also occur~\cite{Manson-2018,Alkauskas-2025}.
Accurate modeling of these effects requires knowledge of carrier capture rates.

In this study, we explore recombination channels and compute capture rates for N$_{\rm C}$ impurities and NV centers in diamond using first-principles calculations. 
We consider capture processes for all charge states that are stable within the band gap: $+1$ through $-1$ for N$_{\rm C}$ and $+1$ through $2-$ for the NV center. 
Carrier capture rates are calculated based on multiphonon emission~\cite{Audrius-2014,Mark-2021}.
For NV$^0$ and NV$^-$ we also consider capture processes mediated by the electronic excited states~\cite{Alkauskas-2016}.
The results allows us to conclude that (1) electron capture on N$_{\rm C}^0$ is fast, and in quantitative agreement with experiment~\cite{Ulbricht-2011}; 
(2) electron capture at N$_{\rm C}^+$ is even faster; and (3) 
for charge transition between NV$^0$ and NV$^-$, the nonradiative processes involving only the electronic ground states are very slow, whereas the ones mediated by the excited states are significantly faster. In particular, hole (electron) capture at NV$^-$ (NV$^0$) proceeds by nonradiative capture into an excited state of NV$^0$ (NV$^-$), followed by radiative decay to the ground state.

The paper is organized as follows. Section~\ref{sec:method} outlines our computational methodology. In Secs.~\ref{sec:N} and \ref{sec:NV}, we present and analyze the calculated atomic structures, formation energies, and capture coefficients for the various charge-state transitions of N$_{\rm C}$ and NV. 
Section~\ref{sec:conclusions} summarizes our findings and conclusions.

\section{Computational Methodology}	\label{sec:method}

Our calculations are based on density functional theory using the Vienna Ab initio Simulation Package (VASP 6.3.1) \cite{kresse-1996, kresse-1996PRB}
with projector augmented wave (PAW) potentials~\cite{kresse-1996PRB, blochl-1994} and a plane-wave basis with an energy cutoff of 420 eV.
We applied the Heyd-Scuseria-Ernzerhof (HSE06) \cite{heyd-2003, heyd-2006} hybrid functional with the default mixing parameter (0.25) and screening parameter (0.2~{\AA}$^{-1}$), which result in an HSE lattice parameter 3.546~{\AA} and band gap of 5.41~eV, in good agreement with the experiments (3.567~{\AA} \cite{Straumanis-1951} and 
5.48~eV~\cite{Cardona-2005}.) 
Point defects are modeled in a 4$\times$4$\times$4 (512-atom) supercell of the conventional 8-atom cubic unit cell; the 
Brillouin zone is sampled at the $\Gamma$ point.
These choices result in accurate electronic properties of N and NV centers \cite{Thiering-2024}.  
The total energy is converged to a $10^{-5}$~eV threshold, and the atoms are allowed to relax while keeping the volume fixed until the forces drop below $10^{-3}$~eV/{\AA}.
 
The formation energy of the relevant point defects in charge state $q$ is 
defined as \cite{Freysoldt-2014}: 
\begin{equation}
	\begin{split}
	E_f^q(E_{\rm F}) = E_{\rm tot}^q - E_{\rm tot}({\rm bulk}) - \sum_{i} n_i \mu_i \\
	+ q(E_{\rm VBM}  + E_{\rm F}) + E_{\rm corr}^q
	\end{split}
		\label{eq:form_E}
\end{equation}
where $ E_{\rm tot}^q$  is the total energy of the supercell containing the defect, and $E_{\rm tot}({\rm bulk}) $ is that of the pristine diamond supercell. $\mu_i$ is the chemical potential of each atom type ($i$ = C, N), while $n_i$ represents the number of atoms of type $i$ added to ($n_i  > 0$) or removed from ($n_i < 0$) the supercell to create the defect. 
$\mu_{\rm C}$ is fixed by equilibrium with bulk diamond, and for purposes of presenting our results we set $\mu_{\rm N}$ equal to the total energy per atom of N$_2$. 
$E_{\rm F}$ is the Fermi level referenced to the valence-band maximum (VBM) and $E_{\rm corr}^q$ is a correction term for the total energy of a supercell containing a charged defect~\cite{Freysoldt-2009}.

To calculate excited-state energies, we use the constrained-occupation $\Delta$-self-consistent-field method \cite{Jones-1989}. 
This approach has been documented to provide excitation energies that agree well with the experimental values~\cite{Mackoit-2019, Aleksei-2023}.

Nonradiative capture processes are studied based on configuration coordinate diagrams (CCDs). 
A CCD maps the potential energy surface as a function of a generalized configuration coordinate $Q$;
the latter is obtained as a linear interpolation of the atomic structures of the initial and final states (before and after charge capture). $\Delta Q$ is the mass-weighted difference between the initial- and final-state structures \cite{Audrius-2014}:
  	\begin{equation}
    (\Delta Q)^2 =  \sum_{J} M_J \left| \mathbf{R}_{J, i} - \mathbf{R}_{J, f}\right|^2 \, 
    \label{eq:delta_q}
  	\end{equation}
where $M_J$ and $\mathbf{R}_J$ are the mass and atomic coordinate of atom $J$. 
The subscripts $i$ and $f$ indicate the initial and final states. 

The nonradiative capture coefficient ($C$) is calculated using the first-principles approach developed by Alkauskas {\it et al.} \cite{Audrius-2014}:
 \begin{equation}
 	\begin{split}
 	C = f \frac{2\pi}{\hbar} g V W_{if}^2  \sum_m w_m \sum_n  \left| \langle \chi_{im} | \hat{Q} - Q_0 | \chi_{fn} \rangle \right|^2 \\
 	 \delta\left( \Delta E + m\hbar \Omega_i - n\hbar \Omega_f \right)
 	\end{split}
 		\label{eq:capture_coeff}
 \end{equation}
where 
$f$ the Sommerfeld parameter, 
$g$ the configurational degeneracy, 
$V$ is the volume of the supercell, 
$W_{if}$ is the electron-phonon coupling matrix element between the initial $i$ and final $f$ states, and
$w_m$ is the thermal occupation factor for the $m$th vibrational level, 
$\chi_{im}$ ($\chi_{fn}$) is the phonon wavefunction of state $i$ ($f$) and vibrational level $m$ ($n$), 
$\hat{Q}$ is the position operator along the configuration coordinate, 
$Q_0 $ is the configuration used for the perturbative expansion~\cite{Mark-2021}, 
$\Delta E$ is the energy difference between states $i$ and $f$, 
and $\Omega_{i, f}$ are the harmonic phonon frequencies of these states.  

Based on the capture coefficient $C$, a capture cross section can be defined as 
\begin{equation}
    \sigma = C/v,
    \label{eq:cap_cross}
\end{equation} 
where $v$=$\sqrt{3 k_B T / m^*}$ is the thermal velocity of the carrier,
in which $k_B$ is the Boltzmann constant and $m^*$ the carrier mass. 
We use the Nonrad code \cite{Mark-2021} to construct the CCDs and compute the capture coefficients or cross sections for the various transitions reported in the paper.

We also study internal transitions, for which the transition occurs between in-gap defect states rather than a carrier being captured from a band edge. 
The corresponding nonradiative rate is
\begin{equation}
    \Gamma_{\rm NR} = C/V, 
    \label{eq:GammaNR}
\end{equation}
where $V$ is the volume of the supercell used in the calculations.

For internal transitions (as described below) we also calculate the {\it radiative} rate~\cite{Mark-2024, Dreyer-2020}:
\begin{equation}
    \Gamma_{\rm R} = \left( \frac{\mathcal{E}_{\mathrm{eff}}}{\mathcal{E}_0} \right)^{2}
    \frac{n_r \mu^{2} (\delta E)^{3}}{3 \pi \epsilon_0 c^{3} \hbar^{4}}.
    \label{eq:GammaR}
\end{equation}
Here $n_r$ represents the refractive index of the host material, which is 2.4 for diamond, and $ \mu $ is the transition dipole matrix element. 
The prefactor $ {\mathcal{E}_{\mathrm{eff}}}/{\mathcal{E}_0} $ denotes the local field effect, which reflects the difference in electric field strength at the defect compared to the bulk material; it is typically close to unity, and we use that value as an approximation \cite{Mark-2024, Razinkovas-2021}. \( \delta E \), finally, denotes the transition energy.

When a charge-state transition level is close to a band edge, thermal emission can occur.
Using detailed balance, the thermal emission rate can be derived based on the nonradiative capture coefficient~\cite{Lang-1974,Wickramaratne-2018}:
\begin{equation}
    e_{n(p)} = C_{n(p)} N_{c(v)} \exp\left(- \frac{\Delta E_{n(p)}}{k_B T}\right).
    \label{eq:thermal_emission}
\end{equation}
Here $C_{n(p)}$ represents the nonradiative capture coefficient for electrons (holes), and $N_{c(v)}$ is the effective density of states of the conduction (valence) band. 
$\Delta E_{n(p)}$ is the charge-state transition level referenced to the CBM (VBM).

\section{Substitutional nitrogen (N$_{\rm C}$)} 
\label{sec:N}

\subsection{Structure and energetics} 
\label{subsec:N_struct_energy}

\begin{figure}[h]
    \centering\includegraphics[scale=0.5, angle=0]{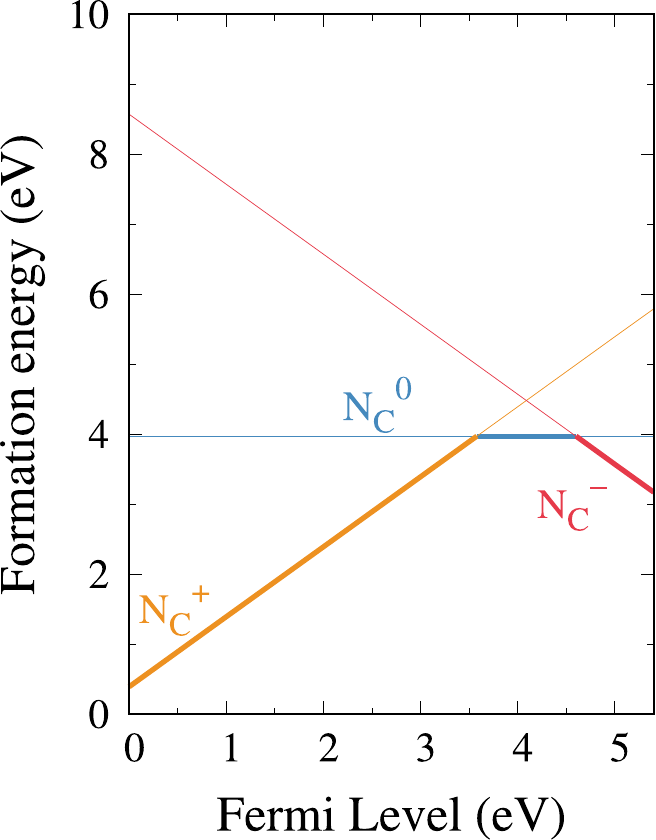}
  	\caption{Formation energy of the substitutional nitrogen (N$_{\rm C}$) as a function of Fermi level referenced to the VBM.}
  	\label{form_E_N}
\end{figure}

\begin{figure}[h]
  \centering\includegraphics[scale=0.22, angle=0]{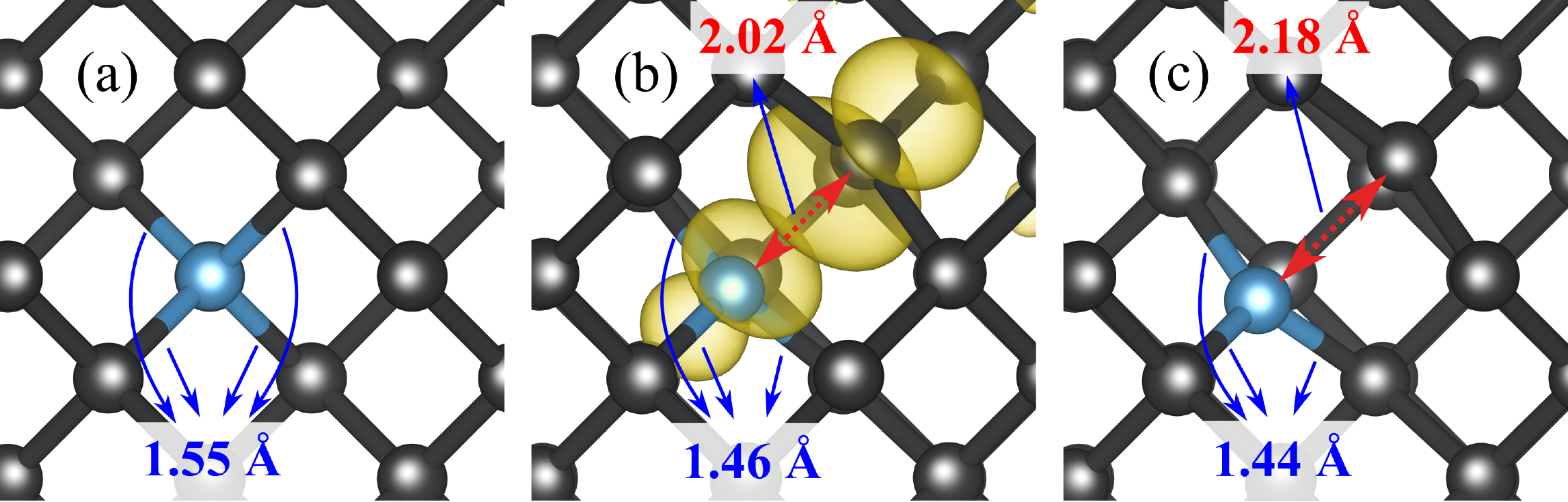}
  \caption{ Ground-state atomic structures of N$_{\rm C}$ for the (a) positive, (b) 
  	neutral, and (c) negative charge states. Carbon and nitrogen atoms are represented by grey and blue spheres.
    In panel (b) the spin density of N$_{\rm C}^0$ is indicated by the yellow isosurface.}
  \label{struc_N}
\end{figure}

Figure~{\ref{form_E_N}} shows formation energies of N$_{\rm C}$ in the positive, neutral, and negative charge states. 
Nitrogen has five valence electrons, while carbon has four. 
In the positive charge state, N$_{\rm C}^+$ occupies the substitutional site with four equivalent N-C bonds, each 1.55~{\AA}\ in length, as seen in Fig.~\ref{struc_N}(a). 
In the neutral charge state, the center undergoes a large lattice relaxation, in which N moves towards the plane of three C neighbors with bond lengths of 1.46~{\AA}.
The N atom assumes more of an sp$^2$ bonding configuration, with two electrons occupying the orbital along the N-C axis, forming a lone pair. 
The fourth bond is effectively broken, consistent with the N-C distance increasing by 
32\% relative to the bulk bond length (1.535 {\AA}), as shown in Fig.~\ref{struc_N}(b). 
The additional electron occupies a 
dangling $p$ orbital on the C atom. This dangling-bond orbital can accommodate an additional electron,
resulting in the negatively charged N$_{\rm C}^-$, in which the N-C distance is stretched 
to 42\% [Fig.~\ref{struc_N}(c)]. 
N$_{\rm C}^0$ and N$_{\rm C}^-$ exhibit \(C_{3v}\) symmetry.

The formation-energy diagram (Fig.{~\ref{form_E_N}}) shows that the $(+/0)$ charge-state transition level occurs at \(E_c - 1.78\)~eV, 
in good agreement with previous theoretical studies (\(E_c - 1.8\)~eV~\cite{Deak-2014}).
Our value also aligns well with temperature-dependent conductivity measurements, which produced
\(E_c - 1.73\)~eV \cite{Farrer-1969, Koppitz-1986}. 
The calculated $(0/-)$ level occurs at \(E_c - 0.84\)~eV, again in agreement with previous calculations (\(E_c - 0.8\)~eV,~\cite{Deak-2014}).
To our knowledge there are no experimental determinations of the (0/$-$) level.

\subsection{Capture processes} \label{subsec:N_capture}

\begin{figure}[h]
  	\centering\includegraphics[scale=0.38, angle=0]{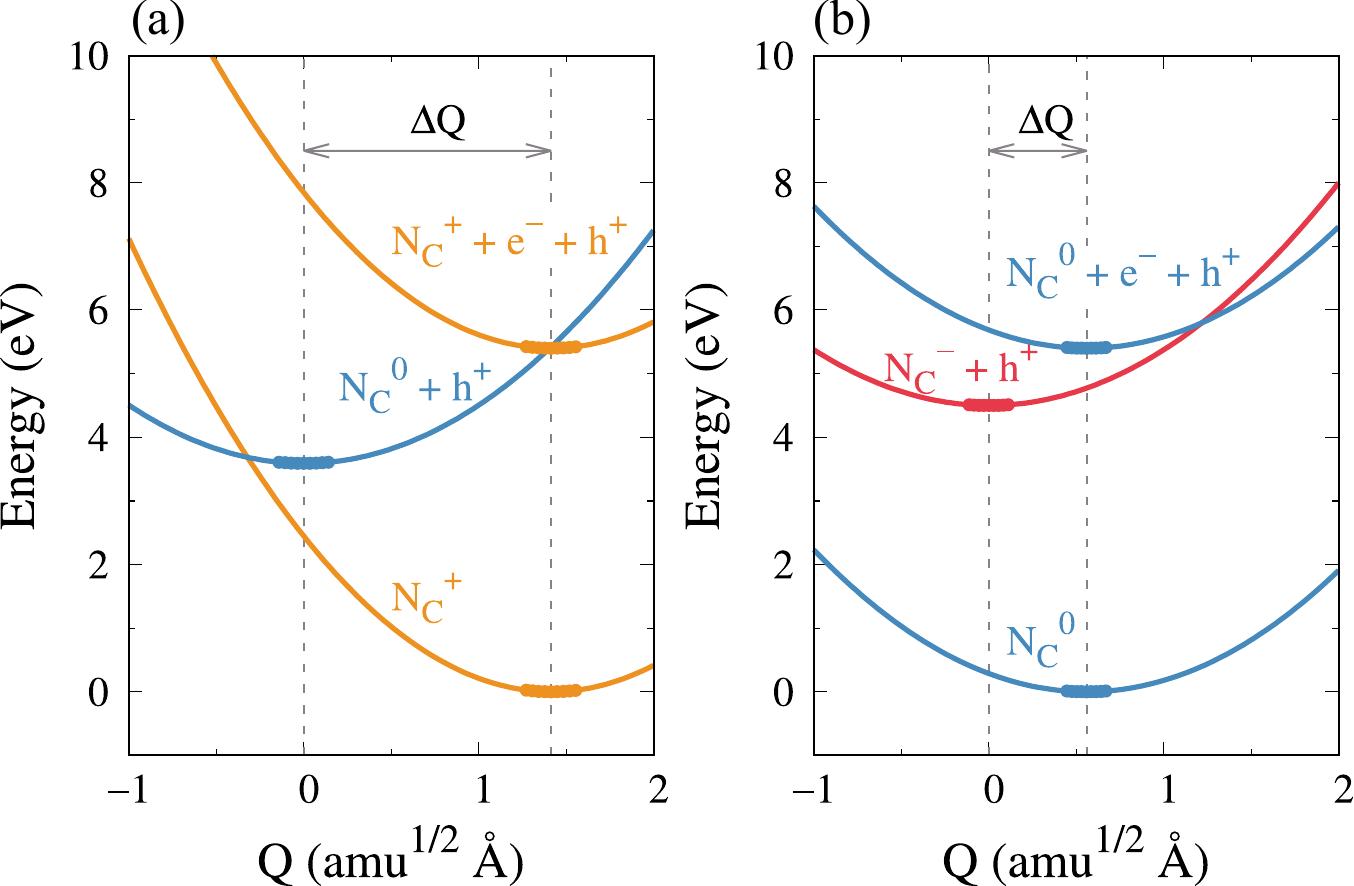}
  	\caption{ Configuration coordinate diagrams for the (a) N$_{\rm C}^+$ $\rightleftharpoons$ N$_{\rm C}^0$ and (b) N$_{\rm C}^0$ $\rightleftharpoons$ N$_{\rm C}^-$ transitions.}
  	\label{ccd_N}
  \end{figure}

The CCDs in Figs.~\ref{ccd_N}(a) and (b) illustrate the nonradiative capture processes for the (+/0) and (0/$-$) transitions.
We first consider capture processes involving the (+/0) level [Fig.~\ref{ccd_N}(a)]. 
In a semiclassical picture of the nonradiative process, the energy barrier defined by the crossing point between the potential energy surfaces determines the rate of the capture process. 
In spite of the fact that the (+/0) level is quite far from both the CBM and the VBM, we find that these barriers are remarkably small, both for capture of an electron from the CBM at N$_{\rm C}^+$ and for capture of a hole from the VBM at N$_{\rm C}^0$, indicating that capture rates will be high.

\begin{figure}[h]
  	\centering\includegraphics[scale=0.5, angle=0]{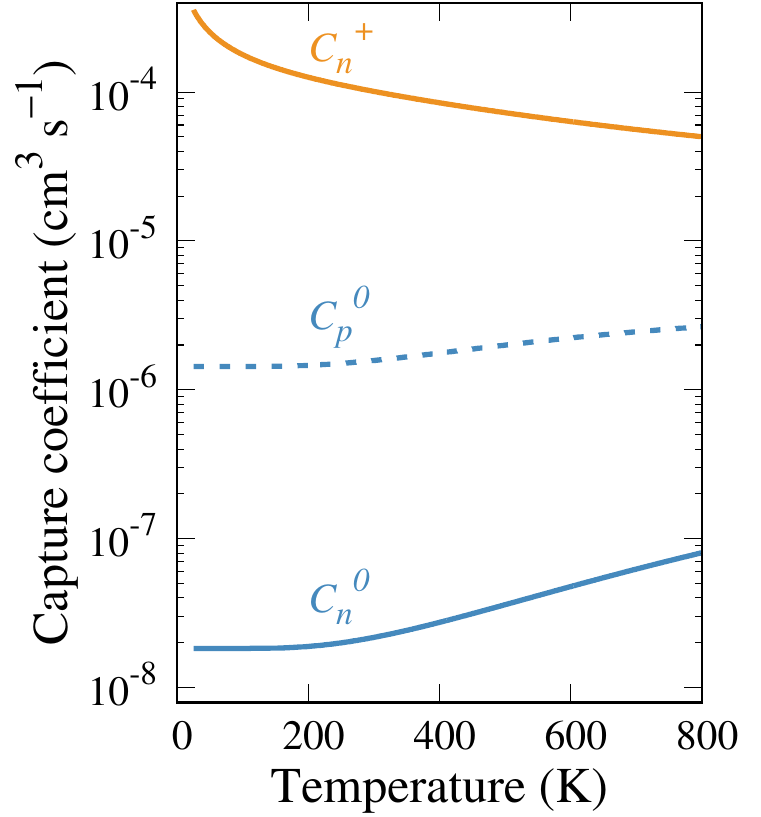}
  	\caption{ Calculated capture coefficients as functions of temperature for N$_{\rm C}$.} 
  	\label{nrc_N}
  \end{figure}

Our full calculations of capture rates are of course quantum-mechanical [Eq.~(\ref{eq:capture_coeff})] \cite{Audrius-2014}, leading to the values that are shown in Fig.~\ref{nrc_N} as a function of temperature.
We use the notation $C_{n(p)}^q$ to indicate electron (hole) capture coefficient at charge state $q$.
In the case of electron capture, the Coulomb attraction between the positively charged N impurity and the negatively charged carrier, which is quantified in the Sommerfeld factor (Sec.~\ref{sec:method}), leads to an enhancement of the capture rate $C_n^+$, particularly at low temperatures.

At room temperature, we find 
$C_n^+$ = 1.0 $\times$ 10$^{-4}$~cm$^3$~s$^{-1}$ 
and $C_p^0$ = 1.6 $\times$ 10$^{-6}$~cm$^3$~s$^{-1}$.
To illustrate the magnitude of these coefficients, we can consider a scenario with N$_{\rm C}^+$ concentration [N$_{\rm C}^+$] $\sim$ 
10$^{14}$~cm$^{-3}$ (as estimated in Ref.~\cite{Ulbricht-2011}).
The rate at which an electron in the conduction band will be captured at any N$_{\rm C}^+$ impurity will be [N$_{\rm C}^+$]$C_n^+$ $\sim$ $10^{10}$~s$^{-1}$, 
i.e., the electron will be captured within 
0.1~nanosecond. 
While this is a high rate, we will see that under conditions of the experiment in Ref.~\cite{Ulbricht-2011}, electron capture at N$_{\rm C}^0$ will be even faster.

We now consider carrier capture at the (0/$-$) level . 
Hole capture at N$_{\rm C}^-$ has an exceptionally large barrier [with the crossing point between the potential energy curves falling outside the range of Fig.~\ref{ccd_N}(b)]. 
As a result, our calculated hole capture coefficient is less than $10^{-30}$~cm$^3$~s$^{-1}$ across the entire temperature range. 
The electron capture process at N$_{\rm C}^0$, on the other hand, occurs with a small barrier, 
leading to a capture coefficient $C_n^0$ = 
2.2 $\times$ $10^{-8}$~cm$^3$~s$^{-1}$ at 300 K [Fig.~\ref{nrc_N}]. 

This result for electron capture can be compared to experiment.
Ulbricht {\it et al.}~\cite{Ulbricht-2011} reported a high capture cross section for electron capture at N$_{\rm C}^0$. 
They excited the sample containing N$_{\rm C}^0$ at various temperatures, measured the conductivity (which is proportional to the excited free-electron concentration $n$) as a function of time, and fit it with an exponentially decaying function $\exp(-t/\tau)$. 
At room temperature, for the sample with total N$_{\rm C}$ concentration of 150~ppm ($3 \times 10^{19}$~cm$^{-3}$), they found $\tau$ in the picosecond timescale. 

Based on our calculated $C_n^0$, we can estimate the rate at which an excited electron is captured at N$_{\rm C}^0$.
Since (as mentioned above) the ionized N$_{\rm C}^+$ concentration in Ref.~\onlinecite{Ulbricht-2011} is only $\sim~10^{14}$~cm$^{-3}$, the N$_{\rm C}^0$ concentration is virtually unchanged from the total N$_{\rm C}$ concentration of 
$3 \times 10^{19}$~cm$^{-3}$.
Therefore, the electron capture rate is [N$_{\rm C}^0$]$C_n^0$ 
$\approx$ $6 \times 10^{11}$~s$^{-1}$, 
corresponding to a lifetime on the order of 
picoseconds, 
consistent with experiment.

\begin{figure}[h!]
    \centering\includegraphics[scale=0.5, angle=0]{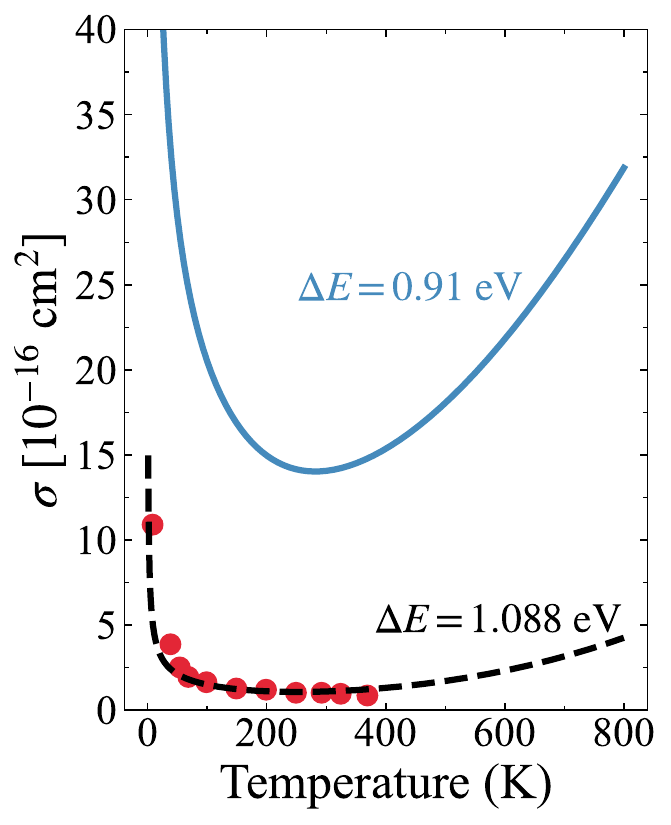}
    \caption{Electron capture cross section at N$_\mathrm{C}^0$ from our calculations (blue curve, using our calculated $\Delta E = 0.91$~eV) and experiments~\cite{Ulbricht-2011} (red circles). 
    The dashed black curve corresponds to our calculations but with $\Delta E$ adjusted to best fit the experimental data, as described in the text.}
    \label{fig:cap_cross}
\end{figure}

Furthermore, Ref.~\onlinecite{Ulbricht-2011} also converted the lifetime $\tau$ to an electron capture cross section $\sigma$ [Eq.~(2) in Ref.~\onlinecite{Ulbricht-2011}] as a function of temperature, which we plot as red circles in Fig.~\ref{fig:cap_cross}. To compare with this, 
we calculated $\sigma$ [Eq.~\eqref{eq:cap_cross}] 
as a function of temperature. 
As shown in Fig.~\ref{fig:cap_cross} (blue curve), our results agree with experiment to 
within about 
an order of magnitude, 
which is gratifying given that our results are obtained fully from first principles.
We suspect that the deviation mainly results from the error bar on the transition energy $\Delta E$.
To test this hypothesis, we allowed $\Delta E$ to vary, finding that a value of 
1.088~eV 
provides an excellent fit to the experimental values (dashed black curve in Fig.~\ref{fig:cap_cross}).
This $\Delta E$ value is only 
0.178~eV 
larger than our first-principles value; a slight underestimate in the calculated $\Delta E$ value is actually expected, since our calculated HSE06 band gap underestimates the experimental gap.

Ulbricht {\it et al.}~\cite{Ulbricht-2011} also mentioned that the thermal emission rate for N$^-$ $\to$ N$^0$ + e$^-$ (i.e. the reverse of the electron capture process) could not play a role in the conversion from N$^-$ to N$^0$, which was found to occur on a nanosecond timescale. 
Our calculated thermal emission rate [Eq.~\eqref{eq:thermal_emission}] at 300~K is 
$e_n$=$1.2 \times 10^{-4}$~s$^{-1}$. This is equivalent to a lifetime of $\sim$8,000~s, 
indeed much larger than the nanosecond timescale. 
Plausible mechanisms could be Auger-Meitner process~{{\cite{Ulbricht-2011}}} or tunneling\cite{Manson-2018, Alkauskas-2025}.

\section{The nitrogen-vacancy (NV) center} \label{sec:NV}

\subsection{Structure and energetics} \label{subsec:NV_struct_energy}

The NV center consists of a carbon vacancy next to a substitutional N atom.
The N atom contributes five valence electrons. Three electrons participate in covalent bonds with neighboring carbon atoms;
the other two, plus the three electrons from C atoms surrounding the vacancy, populate the defect states (Fig.~1 in Ref.~\cite{Thiering-2024}).

\begin{figure}[h]
    \centering\includegraphics[scale=0.5, angle=0]{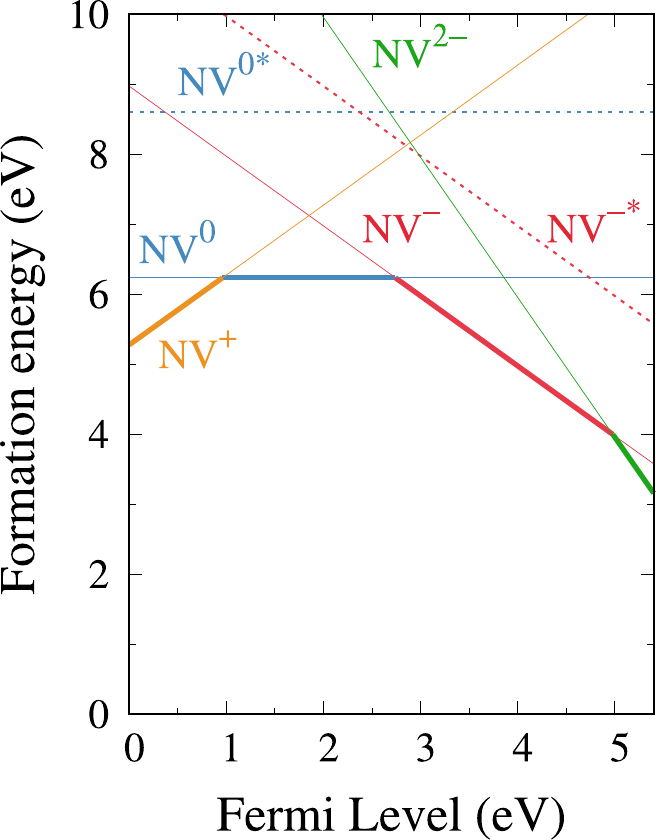}
  	\caption{Formation energy of the nitrogen-vacancy (NV) center as a function of Fermi level referenced to the VBM. Solid lines correspond to electronic ground states; dashed, excited states (labeled by asterisks)}
  	\label{form_E_NV}
\end{figure}

\begin{figure}[h]
 	\centering\includegraphics[scale=0.30, angle=0]{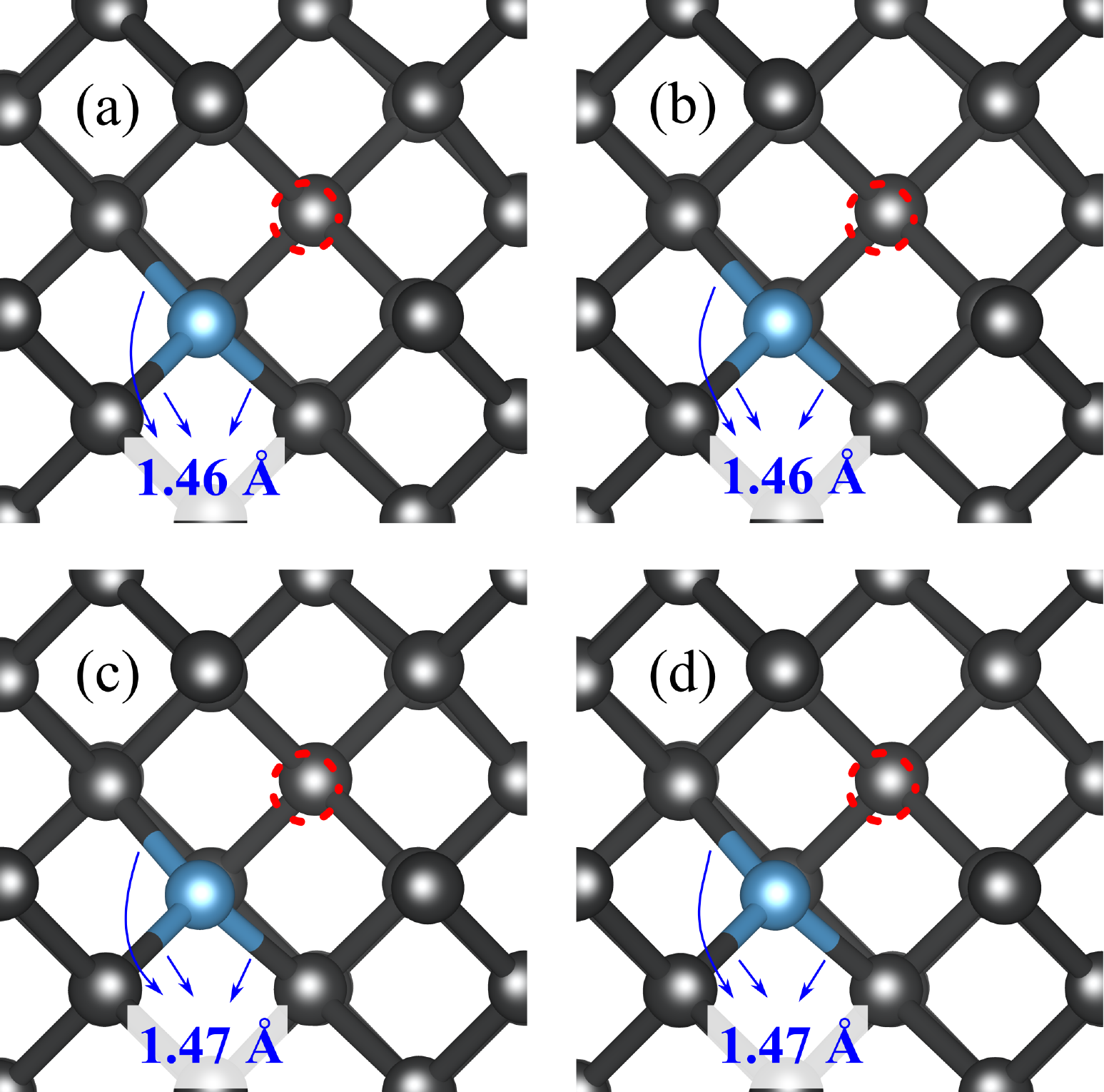}
 	\caption{Ground-state atomic structures of the NV center in the 
    (a) $+$, 
    (b) $0$, 
    (c) $-$, and 
    (d) $2-$ charge states. 
    Nitrogen is blue; carbon, grey; vacancy, dashed red.
    }
 	\label{struc_NV}
 \end{figure}

Four charge states ($+$, $0$, $-$, and $2-$) 
are stable in the gap (Fig.~{\ref{form_E_NV}}).
The relaxed geometries of the ground state structures are shown in Fig.~\ref{struc_NV}. 
The three nitrogen-carbon (N-C) bonds have a value of $\approx$1.46-1.47~{\AA} for all four charge states. 
The trigonal C$_{3v}$ symmetry is 
characterized by three-fold rotational symmetry along the $\langle111\rangle$ axis extending from the vacancy site toward the N atom.  
The NV center can thus occur in four possible orientations.

As seen in Fig.~{\ref{form_E_NV}}, the charge-state transitions for the ground-state configurations occur at 
0.96~eV for ($+/0$), 2.74~eV for ($0/-$), and 4.98~eV for ($-/2-$), all referenced to the VBM. 
Previous calculations (which also used the HSE06 functional) reported 0.9~eV, 2.7~eV, and 4.9~eV~\cite{Deak-2014}.

\begin{figure}[h]
	\centering\includegraphics[scale=0.35, angle=0]{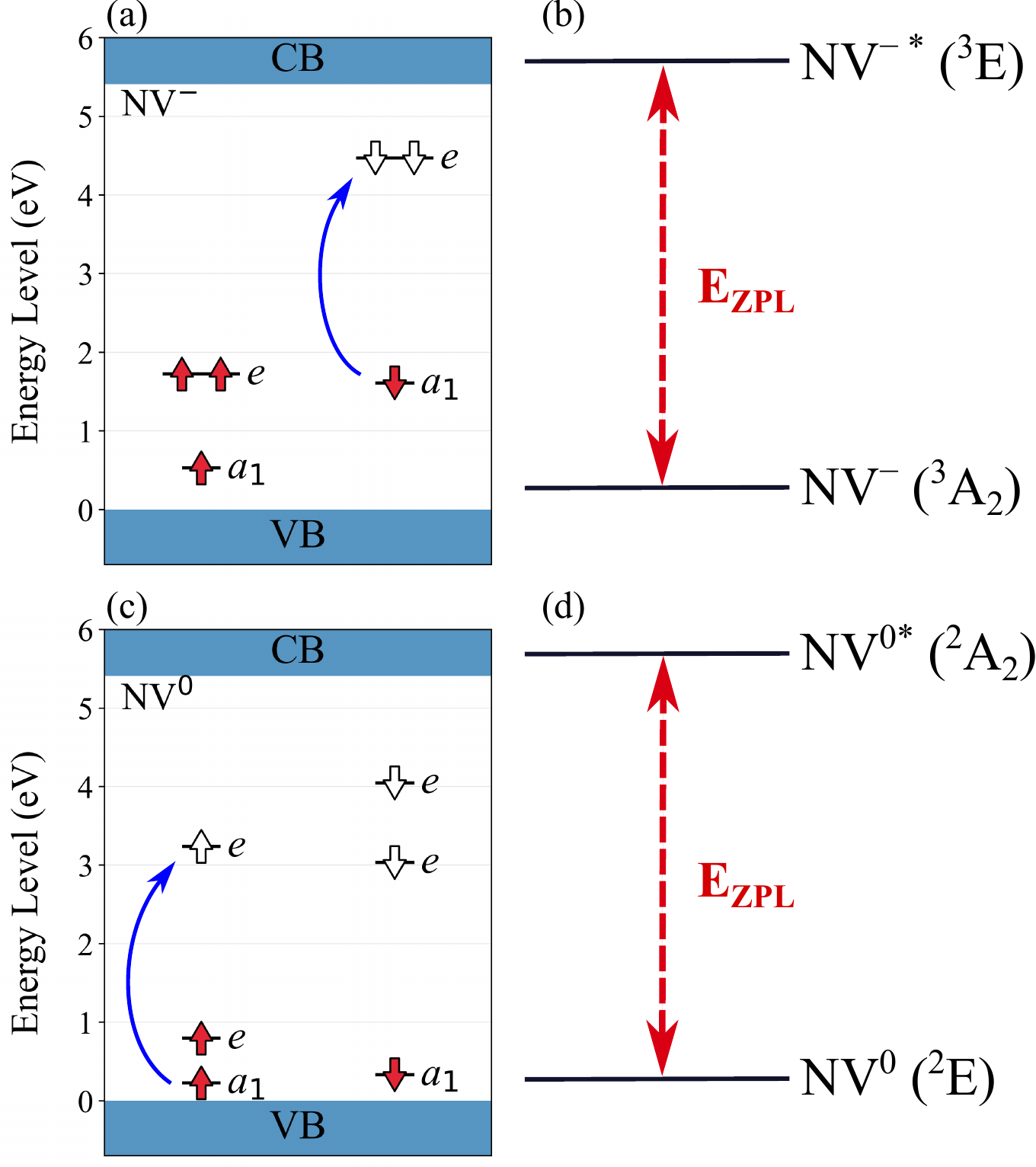}
	\caption{ Kohn-Sham states for the nitrogen-vacancy (NV) center in the (a) negative and (c) neutral charge state. VB and CB label the valence and conduction band, and excitation processes are indicated with curved arrows. The states' 
    symmetries 
    and zero-phonon-line (ZPL) transitions are illustrated in (b) and (d). 
    The illustrated spin configurations are used to calculate the total energies in the present work, where up (down) arrows represent the spin-majority (minority) channel.
    }
	\label{ks_diag_nv}
\end{figure}

Using the orbital notations of the $C_{3v}$ symmetry, 
the Kohn-Sham states in the band gap include an $a_1$ and a doubly degenerate $e$ level [Fig.~\ref{ks_diag_nv}].
For NV$^-$ [Fig.~\ref{ks_diag_nv}(a)], four electrons occupy these levels, resulting in a $^3A_2$ ground state with an 
($a_1a_1ee$) configuration. The first excited state arises from promoting an electron from the $a_1$ to an $e$ orbital, forming the $^3E$ state (labeled as NV$^{-*}$ below) with an ($a_1eee$) configuration. 
(Due to the degeneracy of both $e$ orbitals, promoting the electron to either of them would result in the same total energy.) 
This $^3E$ state is well established both experimentally and theoretically, with a calculated zero-phonon line (ZPL) of 2.00~eV, in good agreement with the experimentally observed 637 nm (1.945 eV) ZPL emission \cite{Doherty-2013}. 

For NV$^0$, three electrons occupy the defect levels, resulting in a $^2E$ ground state with an ($a_1a_1e$) configuration [Fig~\ref{ks_diag_nv}(c)]. 
An electronic excitation from the $a_1$ to an $e$ orbital gives rise to the excited $^2A_2$ (NV$^{0*}$) state with an ($a_1ee$) configuration. 
We obtain the ZPL [Fig.~{\ref{ks_diag_nv}}(d)] in a two-step process. 
First, we calculate the vertical excitation energy from the ground state.
To address the multiconfigurational character of the $^2A_2$ excited state, 
we apply the von Barth scheme~\cite{Barth-1979,Thiering-2024}.
We obtain the energy of the $^2A_2$ state by using Eq.~(12) of Ref.~\onlinecite{Thiering-2024} with both terms 
[using the ($a_1^\uparrow e^\uparrow e^\uparrow$) configuration to calculate the energy of the $^4A_2$ state, where $\uparrow$ labels the spin-majority channel]
calculated in the \emph{ground-state} equilibrium structure, and subtracting the energy of the $^2E$  state; this results in a vertical excitation energy of 2.47~eV. 
We then subtract the Franck-Condon relaxation energy, whose value of 0.18~eV is obtained based on the one-dimensional phonon frequency of the mixed ($a_1^\downarrow e^\uparrow e^\uparrow$) state and $\Delta Q$ between the ground-state equilibrium structure and the approximate structure of the $^2A_2$ state [Eq.~(22) of Ref.~\onlinecite{Thiering-2024}]. 
The resulting ZPL is 2.29~eV, in reasonable agreement with the measured ZPL 2.156~eV~\cite{Davies-1979, Manson-2005}. 

\subsection{Capture processes} \label{subsec:NV_capture}

\subsubsection{Capture into ground states} \label{subsubsec:NV_capture_ground}

\begin{figure*}[htb]
	\centering\includegraphics[scale=0.40, angle=0]{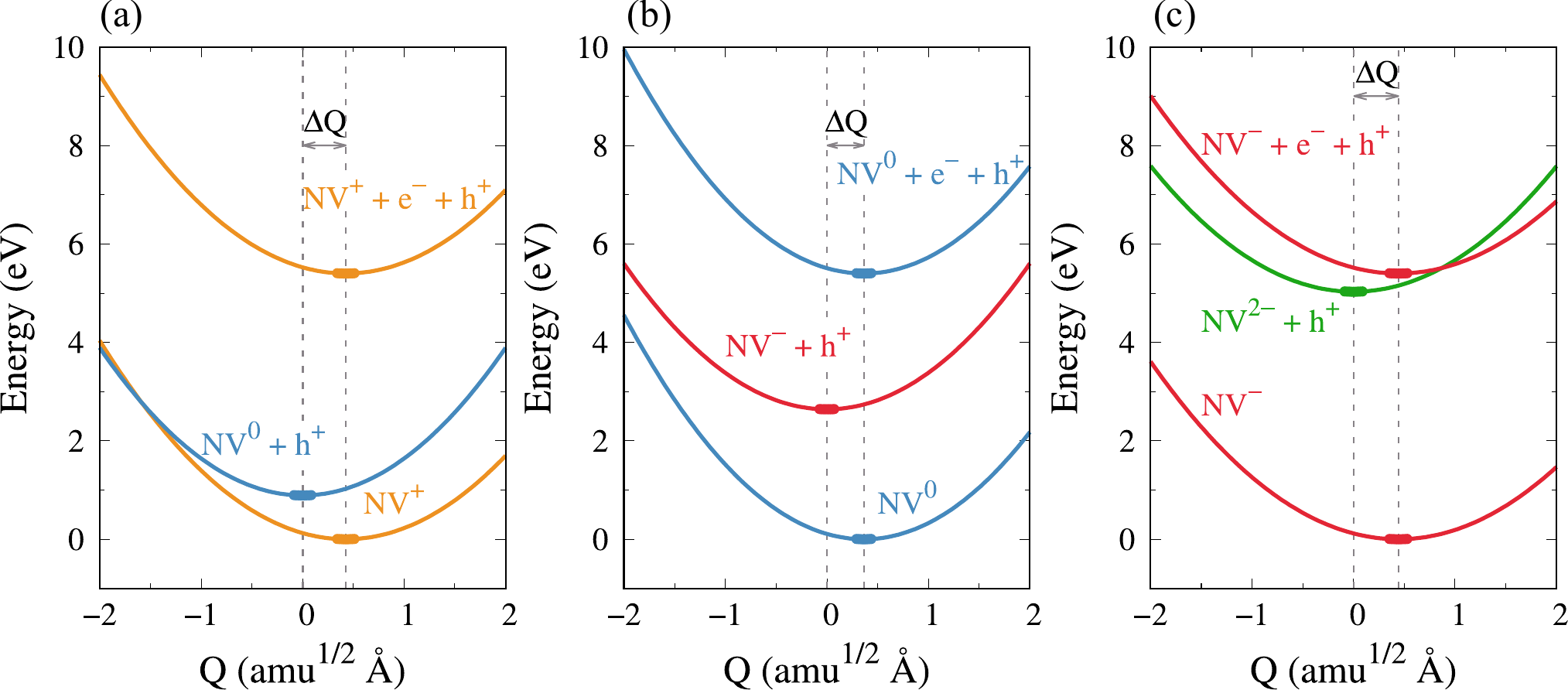}
	\caption{Configuration coordinate diagrams for 
		(a) NV$^+$ $\rightleftharpoons$ NV$^0$, (b) $\rm{NV^0} \rightleftharpoons \rm{NV^-}$, and (c) $\rm{NV^-} \rightleftharpoons \rm{NV^{2-}}$ transitions.}
	\label{ccd_NV_def}
\end{figure*}

\begin{figure}[h]
    \centering
    \includegraphics[scale=.5, angle=0]{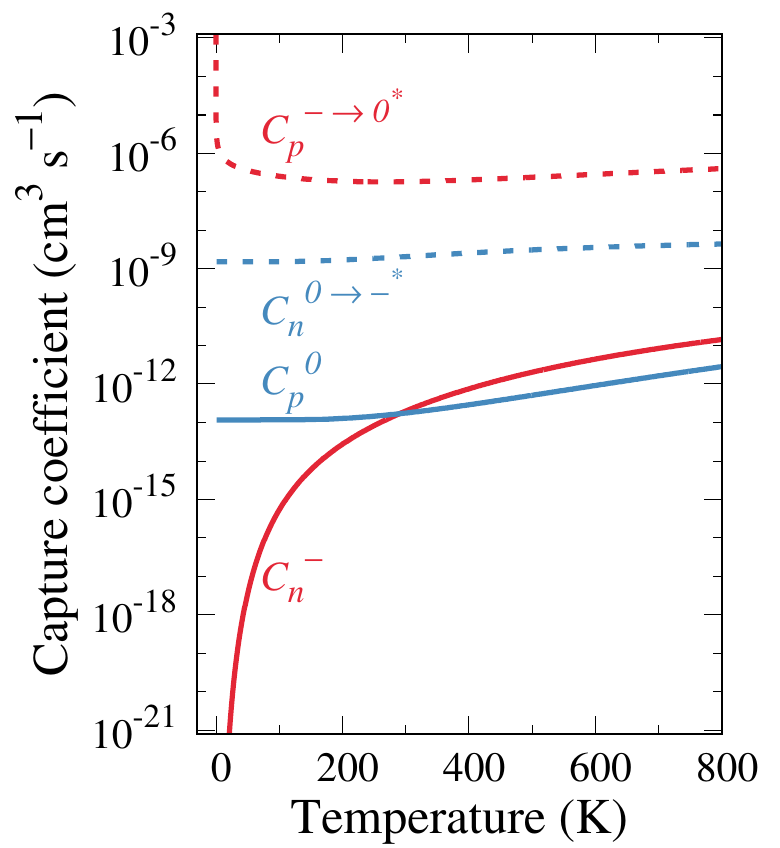}
	\caption{ Calculated capture coefficients as functions of temperature for NV center in the ground (solid curves) and excited (dashed) states.}
	\label{nrc_NV_all}
\end{figure}

Figure~\ref{ccd_NV_def} shows the CCDs for the $\rm{NV^+} \rightleftharpoons \rm{NV^0}$, $\rm{NV^0} \rightleftharpoons \rm{NV^-}$, and $\rm{NV^-} \rightleftharpoons \rm{NV^{2-}}$ transitions. 
For $\rm{NV^0} \rightleftharpoons \rm{NV^-}$, the CCD shows high barriers, resulting in very low capture coefficients ($<$10$^{-28}$~cm$^3$~s$^{-1}$). 
For $\rm{NV^+} \rightleftharpoons \rm{NV^0}$, the electron capture at NV$^+$ has a large energy barrier and therefore a very low capture coefficient of $\sim$10$^{-39}$~cm$^3$~s$^{-1}$, whereas hole capture at NV$^0$ has a significantly lower barrier and therefore a much higher capture coefficient: 
$C_p^0$=$1.8 \times 10^{-13}$~cm$^3$~s$^{-1}$ 
at room temperature (Fig.~\ref{nrc_NV_all}). 
We also calculate the thermal emission rate for NV$^+$ $\to$ NV$^0$ + h$^+$, finding 
$2.4 \times 10^{-10}$~s$^{-1}$ 
at 300~K. 

For the ($-$/2$-$) transition, which occurs within 0.43~eV of the CBM, the electron capture coefficient is 
$C_n^-$=$2.0 \times 10^{-13}$~cm$^3$~s$^{-1}$ 
at 300~K. 
Conversely, hole capture is very slow ($\sim$10$^{-38}$~cm$^3$~s$^{-1}$).

\subsubsection{Capture into excited states} \label{subsubsec:NV_capture_excited}

As discussed in the previous section, nonradiative transitions between NV$^-$ and NV$^0$ are very slow, with capture coefficients $<10^{-28}$~cm$^3$~s$^{-1}$. However, Fig.~{\ref{form_E_NV}} suggests that the transition might be faster when mediated by the NV$^{-*}$ and NV$^{0*}$ excited states. 
Indeed, the charge-state transition level between the neutral charge state and the excited state of the negative charge state (labeled $-*$) is within 0.67~eV of the CBM; and the charge-state transition level between the negative charge state and the excited state of the neutral charge state (labeled 0*) is even closer to the VBM, at 
0.45~eV. 

\begin{figure}[h]
    \centering
    \includegraphics[scale=0.38, angle=0]{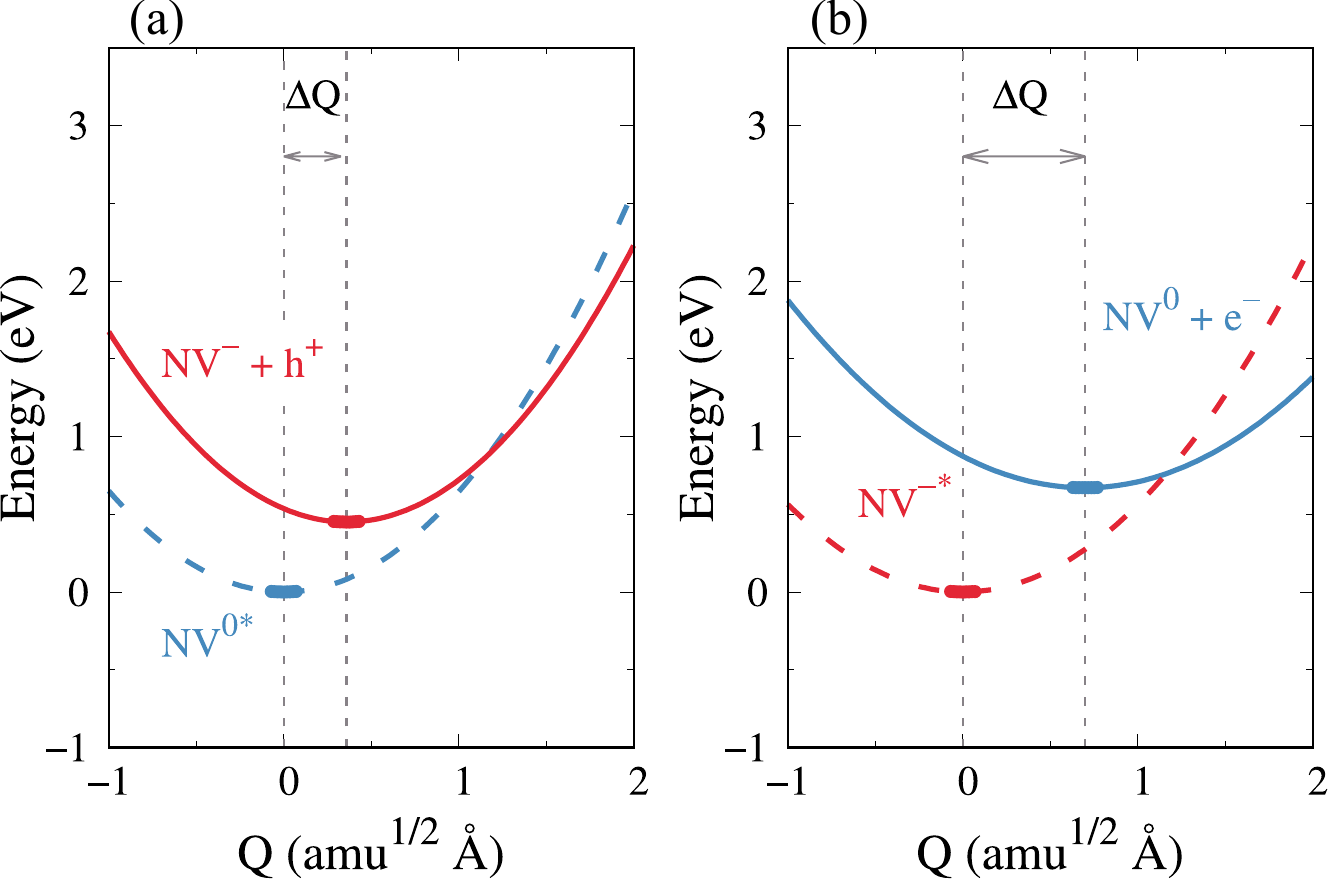}
	\caption{ Configuration coordinate diagrams for charge-state transitions involving excited states: (a) $\rm{NV^-} \rightarrow \rm{NV^{0*}}$, (b) $\rm{NV^0} \rightarrow \rm{NV^{-*}}$. }
	\label{ccd_exc_NV}
\end{figure}

Focusing first on hole capture into NV$^-$, Fig.~\ref{ccd_exc_NV}(a) indeed shows that the barrier height for the NV$^-$ $\rightarrow$ NV$^{0*}$ transition is small, leading to a  hole capture coefficient 
$1.8 \times 10^{-7}$~cm$^3$~s$^{-1}$ 
at 300~K (see Fig.~\ref{nrc_NV_all}), much higher than for NV$^-$ $\rightarrow$ NV$^{0}$.
The increase in $C_p^{- \rightarrow 0*}$ at temperatures below 10~K is due to the Sommerfeld factor~\cite{mark_sommerfeld}, which enhances the attraction of a hole to a negatively charged center.

This process leaves the defect in an excited state; we therefore also need to examine the rate at which the system can relax back to the ground state.
This internal transition can occur either radiatively [rate = $\Gamma_{\rm R}$, Eq.~\eqref{eq:GammaR}] or nonradiatively [rate = $\Gamma_{\rm NR}$, Eq.~\eqref{eq:GammaNR}]. 
The decay process could compete with emitting a carrier from the excited state back to the band edge.
We therefore also assess thermal emission from the excited state,
which can be sizable if the transition levels are close to the band edges.

\begin{figure}[h]
    \centering\includegraphics[scale=0.38, angle=0]{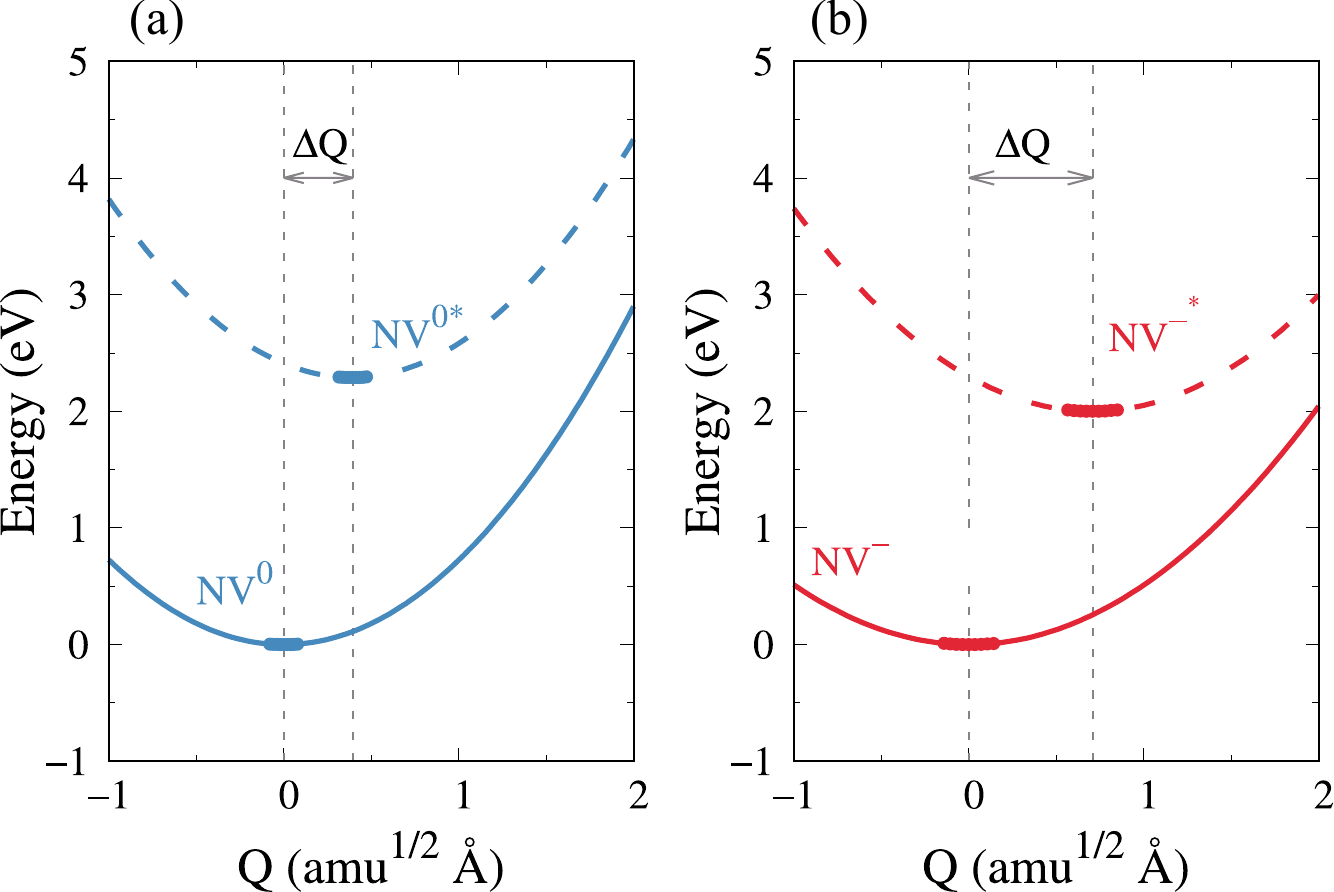}
	\caption{Configuration coordinate diagrams for internal transitions in the NV center: (a) $\rm{NV^{0*}} \rightarrow \rm{NV^{0}}$,  (b) $\rm{NV^{-*}} \rightarrow \rm{NV^{-}}$ .}
	\label{ccd_exc_NV_int}
\end{figure}

For NV$^{0*}$ $\rightarrow$ NV$^0$, the CCD in Fig.~\ref{ccd_exc_NV_int}(a) shows a very large barrier ($>$2~eV), and hence the nonradiative rate is negligible. A radiative transition is more likely, and our calculated radiative rate is 
$5.9 \times 10^{7}$~s$^{-1}$, 
which agrees well with previous HSE06 calculations~\cite{Thiering-2024} and measurements~\cite{Mizuochi-2012, Gatto-2013}.
We note that, even though the $(0^*/-)$ transition level is quite close to the VBM, our calculated thermal emission rate for NV$^{0*}$ $\to$ NV$^-$ + h$^+$ is 
$9.1 \times 10^4$~s$^{-1}$ at 300~K, 
significantly smaller than the radiative rate.  
Therefore, our overall conclusion is that hole capture into NV$^-$ occurs via NV$^-$ $\rightarrow$ NV$^{0*}$ $\rightarrow$ NV$^0$, i.e., by nonradiative capture into NV$^{0*}$, followed by radiative decay to NV$^0$.

The reader may wonder why we focused on the $^2A_2$ excited state of NV$^{0}$, and not on the $^4A_2$ state, which we calculated to lie 0.48~eV above the $^2E$ ground state.
This $^4A_2$ state is actually not relevant for the capture processes discussed in the present work, because the corresponding (0/$-$) level occurs at 2.26~eV above the VBM (compared with 0.45~eV for the $^2A_2$ excited state), implying that hole capture will be slow.

Lozovoi {\it et al.}~\cite{ref:meriles_giant} reported a room-temperature capture cross section $\sigma_h$=3$\times$10$^{-3}$~$\mu$m$^2$=3$\times$10$^{-11}$~cm$^2$ for capture of holes at NV$^-$, based on an experiment in which holes are generated at a source NV center and capture is monitored at a target NV that can be microns away from the source NV.
Our calculated value for $C_p^{- \rightarrow 0^*}$ is $1.8 \times 10^{-7}$~cm$^3$~s$^{-1}$ at 300~K.
Using Eq.~\eqref{eq:cap_cross} with a thermal velocity $1.3 \times 10^7$~cm/s (based on a hole effective mass equal to 0.8 times the free electron mass~\cite{Rauch-1962}), we obtain an estimated value of this cross section of 
1.4$\times$10$^{-14}$~cm$^2$.
(Note that using an effective diffusion velocity, rather than the thermal velocity, would result in a cross section that could be larger by orders of magnitude.)

While this cross section is smaller than the one reported in Ref.~\onlinecite{ref:meriles_giant}, we feel that our calculated capture coefficient is entirely consistent with the hole capture rate (the quantity that is actually observed in Ref.~\onlinecite{ref:meriles_giant}), which is on the order of seconds.
Such a capture rate requires a local hole density on the order of 1$\times 10^{7}$~cm$^{-3}$,
which is entirely plausible given a hole generation rate of 2$\times 10^5$~s$^{-1}$ and the fact that these holes can move with high mobility throughout the sample (with a size of $2 \times 2 \times 0.2$~mm$^3$=0.8$\times 10^{-3}$~cm$^3$) on the time scale of the observed capture process. 
In fact, we feel that the capture cross section reported in Ref.~\onlinecite{ref:meriles_giant} is not directly representative of hole capture at NV$^-$, as it is impacted by the rate at which holes get consumed (by capture at other centers) while traveling between the source and target NVs, as highlighted in recent detailed experiments by the same group~\cite{Monge_arxiv2025}.

Turning now to electron capture into NV$^0$, the barrier height for NV$^0$ $\rightarrow$ NV$^{-*}$ is also modest [Fig.~\ref{ccd_exc_NV}(b)], leading to an electron capture coefficient 
$2.1 \times 10^{-9}$~cm$^3$~s$^{-1}$ 
for NV$^0$ $\rightarrow$ NV$^{-*}$ at 300~K (see Fig.~\ref{nrc_NV_all}).
For the decay to the ground state, NV$^{-*}$ $\rightarrow$ NV$^-$, the CCD shown in Fig.~\ref{ccd_exc_NV_int}(b) again indicates a very large barrier ($>$1~eV), and therefore the corresponding nonradiative rate is negligibly low. 
For the radiative transition we calculate a rate $8.7 \times 10^{7}$~s$^{-1}$.
This value agrees well with previous HSE06 calculations~\cite{Gali-2019, Thiering-2024} and experimental results~\cite{Batalov-2008}. 
Thermal emission could potentially compete with this radiative decay from NV$^{-*}$ to the NV$^-$ ground state, but
we calculated a thermal emission rate for the NV$^{-*}$ $\rightarrow$ NV$^0$ + e$^-$ process of $1.2 \times 10^{-1}$~s$^{-1}$ 
at 300~K, 
much slower than the radiative rate for NV$^{-*}$ $\rightarrow$ NV$^-$. 
The overall conclusion is that electron capture into NV$^0$ occurs via NV$^0$ $\rightarrow$ NV$^{-*}$ $\rightarrow$ NV$^-$, i.e., by nonradiative capture into NV$^{-*}$, followed by radiative decay to NV$^-$.

Finally, we note that we do not consider the excited state of NV$^+$ 
because its (+*/0) charge-state transition transition level lies below the VBM, as estimated from our calculated (+/0) level (Fig.~{\ref{form_E_NV}}) and the ZPL of 1.6~eV calculated in Ref.~{\cite{Karim-2021}}. 

\section{Conclusions} \label{sec:conclusions}

In conclusion, we utilized first-principles calculations based on density functional theory to investigate nonradiative capture processes at N$_{\rm C}$ impurities and NV centers, as well as radiative decay and thermal emission processes. 

For the N$_{\rm C}$ impurity, we find high capture coefficients for the electron and hole captures involving the (+/0) transition. We also find a high capture coefficient for electron capture to N$_{\rm C}^0$, which agrees well with the experimentally measured capture cross section.

For the NV center, most capture coefficients associated with ground-state transitions (including NV$^0$ $\rightleftharpoons$ NV$^-$) are extremely low. We find that transitions mediated by excited states are significantly faster.
In particular, we find that NV$^0$ $\to$ NV$^{-*}$ $\to$ NV$^-$ (via nonradiative charge capture followed by radiative decay) is much more likely to occur than a direct NV$^0$ $\to$ NV$^-$ transition.
Similarly, NV$^-$ $\to$ NV$^{0*}$ $\to$ NV$^0$ is the likely pathway for hole capture at NV$^-$.

\section*{Acknowledgments}
We thank G. Thiering and Y. Chen for fruitful discussions. 
C.K.V. and J.K.N. were supported by the U.S. Department of Energy (DOE), Office of Science (SC), National Quantum Information Science Research Centers, Co-design Center for Quantum Advantage (C2QA) under contract number DE-SC0012704. 
J.K.N.’s work on excited states of the NV center was supported by the National Science Foundation under grant number OSI-2427159. 
M.E.T.'s work on the methodology of nonradiative recombination was initially supported by the University of California Santa Barbara NSF Quantum Foundry funded via the
Q-AMASE-i program under Award No. DMR-1906325; and subsequently by the
Office of Naval Research through the Naval Research Laboratory's Basic Research Program.
The work used computing resources provided by the National Energy Research Scientific Computing Center (NERSC), a User Facility supported by the DOE SC under Contract No. DE-AC02-05CH11231 using NERSC award BES-ERCAP0021021 and BES-ERCAP0028497.

\section*{Data Availability}
The data that support the findings of this article are openly available~{\cite{ref:nomad}}.

\bibliography{reference_NV}
\end{document}